\documentclass[a4paper,twoside]{article}
\usepackage{cite}
\usepackage{amsmath,amssymb,amsfonts}
\usepackage{algorithmic}
\usepackage{graphicx}
\usepackage{textcomp}
\usepackage{xcolor}
\usepackage{epsfig}
\usepackage{subcaption}
\usepackage{calc}
\usepackage{amssymb}
\usepackage{amstext}
\usepackage{amsmath}
\usepackage{amsthm}
\usepackage{multicol}
\usepackage{pslatex}
\usepackage{apalike}
\usepackage{SCITEPRESS}     
\usepackage{flushend}

\begin{document}
\title{Practical Hash-Based Anonymity for MAC Addresses}
\author{\authorname{Junade Ali\sup{1}\orcidAuthor{0000-0002-0180-070X}, Vladimir Dyo\sup{2}\orcidAuthor{0000-0003-4747-0870} }
\affiliation{\sup{1}Cloudflare Inc, London, UK}
\affiliation{\sup{2}University of Bedfordshire, Luton, UK}
\email{junade@cloudflare.com, vladimir.dyo@beds.ac.uk}
}


\abstract{
Given that a MAC address can uniquely identify a person or a vehicle, continuous tracking over a large geographical scale has raised serious privacy concerns amongst governments and the general public. 
Prior work has demonstrated that simple hash-based approaches to anonymization can be easily inverted due to the small search space of MAC addresses. 
In particular, it is possible to represent the entire allocated MAC address space in $39$ bits and that frequency-based attacks allow for $50\%$ of MAC addresses to be enumerated in $31$ bits.
We present a practical approach to MAC address anonymization using both computationally expensive hash functions and truncating the resulting hashes to allow for $k$-anonymity. 
We provide an expression for computing the percentage of expected collisions, demonstrating that for digests of $24$ bits it is possible to store up to $168,617$ MAC addresses with the rate of  collisions less than $1\%$. 
We experimentally demonstrate that a rate of collision of $1\%$ or less can be achieved by storing data sets of $100$ MAC addresses in $13$ bits, $1,000$ MAC addresses in $17$ bits and $10,000$ MAC addresses in $20$ bits.
}
\onecolumn \maketitle \normalsize \setcounter{footnote}{0} \vfill

\section{Introduction}
The ubiquitous use of Wi-Fi and Bluetooth among drivers, passengers and pedestrians  has enabled   monitoring  transportation flows on a massive scale through deployments of wireless scanners along the roads that passively scan, track and analyze nearby wireless devices. 
In particular, wireless scanner equipment has been deployed for monitoring pedestrian and cyclist journey time \cite{abedi2015assessment}, monitoring road conditions in real-time \cite{abberley2017modelling} and passenger movements throughout train stations \cite{fearn_2019}.  
Given the fact that a MAC address can uniquely identify a person or a vehicle, continuous tracking over a large geographical scale has raised serious privacy concerns amongst governments and the general public \cite{minch2015location,fearn_2019}. 

The current practice adopted by many operators and dataset providers is to anonymize MAC addresses  using hashing algorithms such as SHA256. 
However \cite{marx2018hashing} notes that due to a small pre-image space for MAC addresses, all SHA256 hashed MAC addresses could be recovered in 13 minutes 22 seconds.
Despite MAC addresses containing a 48-bit search space, with only 0.1\% of OUI manufacturer prefixes for MAC addresses being allocated \cite{fuxjaeger2016towards} it is possible to represent all MAC addresses in a digest of 39 bits.  
Furthermore, due to the unequal usage of various OUI prefixes \cite{demir2014analysing}, it is possible to represent 50\% coverage in 31 bits, 90\% coverage in 33 bits and 99\% coverage in 34 bits respectively.
Hence, \cite{fuxjaeger2016towards} proposes simply removing the OUI manufacturer prefix from MAC address before storage. 
However, \cite{martin2016decomposition} finds that contiguous address blocks are allocated to specific devices meaning the residual NIC suffix of the MAC address can similarly be used for identification attacks. 
\cite{demir2017pitfalls} contend that it is difficult to control the anonymity set size when using approximations of the Birthday Paradox.
As shown in our analysis, through  reformulating a Birthday Paradox in \cite{demir2017pitfalls}, we find that the  probability of there being at least one collision exceeds $0.5$ when the number of MAC addresses in the dataset exceeds $4,822$ and the hash is stored in a digest of 24 bits or less.

The evaluation of MAC schemes has been mostly limited to analysing the  "at least one collision" semantics, representing an all-or-nothing approach. 
In this paper we argue that this metric alone is not sufficient for a decision maker to choose an anonymizing solution as 
there is a large class of applications, such as popular route analysis or building a driver's profile, where a small fraction of hash collisions is acceptable as long as does not affect large-scale patterns. 
In these situations, a decision maker is often concerned about minimising the rate of collisions having accepted the fact that the collisions are inevitable. 
In other words, for these applications, a more practical question is to select an anonymizing solution that provides an upper bound on the fraction of MAC addresses that have collisions rather than minimise the probability of them happening  at all.

In this paper we propose a practical approach to anonymization which seeks to explore the trade-off between the fraction of collisions and the hashing mechanism. 
We first develop an expression for calculating the overall rate of collisions  given the number of MAC addresses and a  digest size, instead of merely calculating the approximate probability of being at least one collision. 
We demonstrate that with a database of up to $168,617$ unique MAC addresses and a digest size of up to $24$ bits, the proportion of MAC addresses that have  collisions remains below 1\%.
The proposed approach builds upon recent work on $k$-anonymity concept \cite{li2019protocols,thomas2019protecting}, whereby hashes are truncated such that queries can be made to $k$-anonymous buckets for reducing information leak when searching for data. We experimentally verify this approach by randomly selecting MAC Addresses from a search space of $8.4$ million MAC addresses generated by a RedHat script for virtual machines.

By considering an acceptable probability of address collisions (instead of the acceptable probability of one collision), we provide an improved model for calculating how many bits a hash digest should be truncated for anonymization. 
We demonstrate such calculations are appropriate using empirical experimentation. 
Furthermore, through utilising computationally expensive key derivation functions \cite{thomas2019protecting} instead of simple hash algorithms, we are able to prevent background knowledge attacks documented in \cite{demir2014analysing} and \cite{martin2016decomposition} whilst requiring greater effort to identify the candidate of pre-image MAC addresses.
For a probability of collision less than $1\%$, we find truncating the hash digest to $17$ bits is suitable for $1,000$ MAC addresses and truncation to $20$ bits is suitable for $10,000$ MAC addresses. 
For up to $168,617$ MAC addresses, our approach provides for the number of bits in the digest to be less than or equal to the number of bits exposed in the \cite{fuxjaeger2016towards} approach. 
Unlike \cite{demir2014analysing}, our approach has no dependency on withholding a secret key and no post-collection aggregation to produce a single identifier.

\section{Related Work}
Traditionally, the vehicles have been tracked using Automatic Number Plate Recognition (ANPR) systems, which require expensive camera infrastructure and computationally intensive image recognition systems \cite{blogg2010travel}.  
The traffic volume can also be measured by inductive loops or magnetic sensors built-into the road \cite{vladan2016magnetic}, however the technology is limited to counting and classifying  vehicles and lacks vehicular identification capability. 
The ubiquitous adoption of Bluetooth and Wi-Fi devices among drivers, pedestrians and vehicles themselves have enabled low cost  tracking and monitoring pedestrian and motor traffic on a massive scale. 
The data has been used for pedestrian and cyclist journey time \cite{abedi2015assessment}, modelling road congestion alongside tracking real-time road traffic conditions \cite{abberley2017modelling} and providing journey time and usage data for public transport \cite{hidayata2018time}. 
Real world deployments of such technology include roads in the Highways England network, whereby digital signage indicates journey time based on capturing Bluetooth data from point-to-point. 
One engineering firm, Clearview Intelligence, claims Bluetooth-based vehicular tracking systems are approximately 10\% of the cost of equivalent ANPR solutions \cite{clearview}. Research for Transport Scotland found Bluetooth based sensing was very cost effective due to avoiding digging costs associated with induction loop counters \cite{cragg2013bluetooth}. The costs of Bluetooth data collection are significantly cheaper than other data collection methods \cite{blogg2010travel}, as such these systems are likely to remain as an effective way of measuring transportation flows. 

At the same time, massive deployments of such technology have raised privacy concerns; \cite{minch2015location} notes of a deployment in 2012 where a company called Renew trialled installation of 12 recycling bins in the City of London with the capability of capturing wireless MAC address information; on a single day on the 6th July, 946,016 presences were detected, which included 106,629 unique MAC addresses. 
The company even proposed installing such sensors in restrooms, to allow the gender of the user to be inferred. 
By August 2013, the City of London Corporation had requested the trial be halted although the company claimed they only captured "extremely limited, encrypted, aggregated, and anonymized data."

Similar privacy concerns have appeared more recently, concerns were raised in May 2019 where Transport for London (TfL) announced they would collect MAC address data to understand how passengers move throughout stations \cite{fearn_2019}. Whilst TfL claimed the MAC addresses were hashed (with a salt), privacy advocates argued the approach was not suitable as there was still a need to link the same MAC address across observations.

Although hash-based anonymization schemes have been proposed before, its pitfalls are well documented in \cite{demir2014analysing}. It is shown that due to a small address space, it is possible to reverse hashing of a MAC address with simple brute force attacks in a relatively small time frame which the authors identify as a matter of minutes. The first 24 bits of a MAC address represent the OUI prefix allocated to the vendor, with only a small space being allocated and disparity in usage allowing for the search space to be reduced. The authors compare the hashing approaches used by 15 vendors and find whilst the majority use simple hash based approaches, in two instances random "salt" values are added prior to hashing. The authors propose a hash-chain approach whereby the next security key is derived from a hash of the previous key, when the previous entry is discarded it is not possible to identify the key. Unfortunately, this solution still causes there to be a key in the system, relying on the fact the key is stored in a more secure form than the data itself - an improved solution should provide for anonymization without any such key. Such a chaining approach also raises problems for storage in a decentralised environment.

Such salts do not function like salts in password hashing algorithms, in password hashing algorithms unique salts are stored alongside the passwords and in the event of a data breach, it means dictionaries of passwords need to be computed for every salt value, i.e. unique to each stored password, instead of being able to compare pre-computed hashes with every value in the database \cite{ertaul2016implementation}. In this context, the salt represents more of a key, some vendors may use the same "salt" for all captures over the period of a day, allowing previous records to be found for comparison over that time period. If the "salt" is disclosed, all records can be de-anonymized over the period of time that key is used. \cite{demir2014analysing} concludes by proposing an improved trade-off between searching historical records and securing data, by using an iterative key derived through insertions as the salt of the hash function. Unfortunately this still poses a number of challenges; whilst it provides greater security of a storage system, the key can still be obtained if the server computing the statistics is compromised. Furthermore, the chained approach introduces a search difficulty of $O(n)$ whereas databases can be optimised using tree structures with B-trees allowing search in $O(log\, n)$ time.

\cite{fuxjaeger2016towards} notes hashing algorithms are also unsuitable: "the claims concerning non-reversibility have already been shown to be misleading since the actually allocated part of the 48bit MAC address space can be reverse-mapped in a brute-force way with low computing power." The authors propose that the MAC address is truncated to provide anonymity by removing the OUI (the manufacturer prefix) prior to hashing. Unfortunately, this approach does not appear to be a suitable anonymization approach; as \cite{demir2014analysing} notes, only $0.1\%$ of prefixes are allocated (when excluded brute force space can be reduced $1024$x) and the OUIs are not uniformly distributed with some manufacturer used dramatically more than others. The experimental results in \cite{fuxjaeger2016towards} by failing to find any collisions in their dataset using this model.

\cite{demir2017pitfalls} notes that limits exists to simple implementations of hashing algorithms in a variety of applications and consider the use-case of one-to-many anonymization of MAC addresses. \cite{marx2018hashing} notes that pre-image space is bounded for a lot of personally identifiable information types, in the case of MAC addresses by the first 24 bits being the OUI, reducing the search space to $4.1 \cdot 10^{11}$. With the ability to calculate 6 Giga MD5 hashes and 844 Mega SHA-256 hashes per second the authors are able to recover $100\%$ of 1 million hashes in 4 minutes 1 second for MD5 and 13 minutes 22 seconds for SHA-256 respectively. This work demonstrates that even when searching the entire OUI space, it is not sufficient to use relatively computationally inexpensive hashing functions as adversaries will have significant capacity.

It is worth noting that both Android and Apple iOS operating systems allow for devices in a disassociated state to randomise the MAC addresses used for active scans, allowing for a degree of anonymity until they connect to an access point. 
However,  the results of a wide-scale study \cite{martin2017study}, indicate  that adoption rates of such technology are surprisingly low and   re-identification attacks can be used in 96\% of Android phones. The growth in Bluetooth hands-free equipment and headphones can further reduce the effectiveness of anonymization technology as the device will be in a paired state. In road traffic situations, the presence of Bluetooth hands-free devices, GPS devices and Bluetooth systems inbuilt into vehicle provides for other devices to be used for tracking purposes, even where mobile phones provide a degree of anonymization.

Recently, a number of advancements have appeared in the realm of identification of breached passwords in an anonymized way. \cite{li2019protocols} provides an overview of protocols for {\it compromised credential checking} (C3) services and a mechanism for empirically evaluating such protocols. Such protocols are based off the {\it HIBP} protocol whereby passwords are bucketed based on a hash prefix, allowing for $k$-anonymous verification of a breached password as  implemented in \cite{ali2017mechanism}. This bucket is then served to the user to identify if the password is breached or not. Other protocols utilising two-way guessing games are also described, however do require additional two-way steps in the protocol to achieve such communication. The optimal approach presented is known as IDB (identifier based bucketization) where the prefix of the bucket used is based on a different identifier, like the hash of a username. The authors provide formalisation of such C3 protocols, discussion on threat models and empirical analysis of such protocols. \cite{thomas2019protecting} adds additional layers to such an approach, using computationally expensive hashing and private set intersection alongside $k$-anonymity.


\section{Threat Model}

\begin{figure}[]
\centering
        \includegraphics[totalheight=1.2cm]{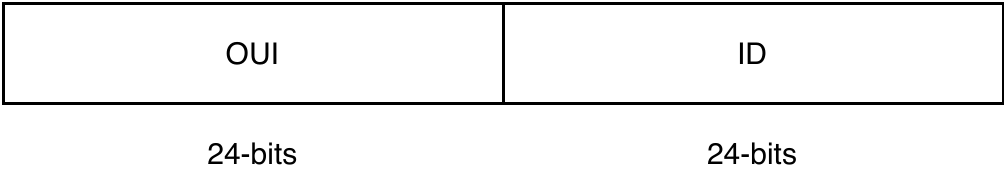}
    \caption{Structure of a MAC Address}
    \label{fig:MACaddress}
\end{figure}

The MAC address is a unique identifier for Wi-Fi, Ethernet and Bluetooth devices and is defined by  IEEE 802 standard.  
The 24 most significant bits of a MAC address form an Organizationally Unique Identifier (OUI) prefix which is registered with the IEEE. The remaining 24 bits represent the Network Interface Controller ID, which uniquely identifies a particular device within the OUI prefix, as shown in Figure \ref{fig:MACaddress}.

The MAC address search space can crudely be considered to be 48 bits corresponding to $2^{48}$ combinations. 
However as \cite{demir2014analysing} notes, given only $0.1\%$ of OUI prefixes are allocated, the effective search space is reduced to $2^{24} \times (2^{24} \times 0.001)$ combinations, which can be expressed in at most $\left\lceil log_2 2^{24} \times (2^{24} \times 0.001) \right\rceil = 39$ bits.

Furthermore, guesswork approaches allow for either prioritisation or limiting the search space further based on the most popular hardware manufacturers. \cite{demir2014analysing} found in an experimental dataset that 50\% coverage was obtained with 87 OUI prefixes, 90\% coverage with 361 prefixes and 709 prefixes covering 99\%. We can accordingly calculate that 50\% coverage can be represented in $\left\lceil log_2 2^{24} \times 87 \right\rceil = 31$ bits, 90\% coverage in 33 bits and 99\% coverage in 34 bits respectively.

This demonstrates that due to the limited entropy provided by the OUI prefix of a MAC address, anonymity is not possible for all records where a MAC address is hashed with a digest size of $39$ bits or more. Experimental data from \cite{demir2014analysing} accordingly demonstrates that frequency-based attacks become possible on 50\% or more of the dataset when the hash is truncated to 31 bits or more.

This attack can also be inverted, due to the small search space when the OUI is removed  as the  described   in \cite{fuxjaeger2016towards}. 
Instead of seeking to identify an individual MAC address from a set, it also allows for data to extract devices of a specific type. 
This approach may be combined with the work in \cite{martin2016decomposition} where contiguous address blocks being allocated to specific devices allows for fine-grained mapping of the devices in use - allowing the device type to be identified.
This reaches to the fundamental dogma of anonymization of MAC address information; attacks may be be deductive in nature, where seek to identify the existence of a MAC address from a given dataset or inductive, which seek to extract enriched data from the dataset, such as in \cite{martin2016decomposition}. 
We therefore must consider these two threats in our anonymity model.

\section{Anonymity Model}

\subsection{Choice of Hashing Algorithm}

It is essential to ensure that when MAC addresses are placed to anonymized buckets, they are unrelated to other records contained into those buckets. Whilst \cite{fuxjaeger2016towards} proposes simply removing the OUI manufacturer prefix from MAC address before storage thus reducing the search space to 24 bits, such an approach is still vulnerable to individuation attacks given \cite{martin2016decomposition} finds that contiguous address blocks are allocated to specific devices meaning the residual NIC suffix of the MAC address can similarly be used for identification attacks and as \cite{demir2014analysing} notes that as only $0.1\%$ of OUI prefixes are allocated.

Fortunately, hash functions counter this issue. The Avalanche effect means a small change in the input results in a completely different output \cite{ali2017mechanism}, this has the effect of distributing devices with the same OUI prefix throughout the dataset of resulting hashes (and contiguous NIC blocks).

\cite{marx2018hashing} notes that due to the small pre-image space of MAC addresses, it is possible to brute force such values in a matter of minutes when using the SHA256 algorithm. To compliment our anonymity approach and deter mapping of anonymized hashes with possible values, we elect to use computationally expensive hashing functions to increase the computational complexity required for brute force attacks, as used with passwords in \cite{thomas2019protecting}.

Key derivation functions like Argon2, PBKDF2, BCrypt or SCrypt may be used with a configurable work factor and a salt input. Such that the resulting digests can be used for simple comparisons of the resulting digest, we propose that the salt is used as a static value which acts as a secret; \cite{alimpia2018enhanced} provides more information on how such a configuration can be used with BCrypt in Expensive Key Setup to generate Message Authentication Codes. Such a salt can be rotated as desired, \cite{demir2014analysing} provides an overview of upper-bound retention time by some manufacturers which may be relevant to determining key rotation policies. By using a computationally expensive key derivation function instead of a more simplistic hash, data is protected by means of both a secret and (in the event the key is compromised) exhaustive search becomes more difficult dependant on a work factor. Deployments where a user identifier is available (such as a registration plate number or an email address) provide for additional entropy to be added into the salt value where applicable.

As noted in \cite{cragg2013bluetooth}, it is critical that hashing is performed at the time of detection prior to the MAC Address being recorded. This approach means that there is no post-processing work required to form a hashed identifier for the MAC Addresses as the resulting digest is consistent. This provides the additional benefit that the work function can be set to a high value to deter brute force attacks without the need to compute hashes in bulk for post collection analysis, as required by \cite{demir2014analysing}.

The maximum work factor of a hash function is limited by the resources available compute the hash and the traffic rate. \cite{blocki2018economics} describes that it is increasingly trivial for an attacker to gain vast amounts of computational resources for cracking hashes and accordingly recommends use of memory hard functions (MHFs) such as SCrypt or Argon2 to reduce the damage of offline attacks (in contrast to non-memory hard functions such as BCrypt or PBKDF2). The Argon2 hash algorithm won the Password Hashing Competition (PHC) run from 2013 to 2015 \cite{wetzels2016open} and is described in \cite{biryukov2016argon2}. Argon2 exists in two variants, Argon2i (optimised for resisting side-channel attacks) and Argon2d (instead optimised for resisting GPU accelerated cracking attacks). Whilst side-channel attacks are a risk for authentication systems, such threats do not exist ordinarily for journey time monitoring systems with no user interaction. Given the limited entropy of MAC Addresses poses a vastly greater risk, we accordingly select the Argon2d algorithm.

\subsection{Bucketization of Hashes}

We seek to devise an approach whereby hashes are placed into anonymous buckets, with an acceptable upper bound on acceptable collision rate. 
Whilst there is a low probability a MAC address is unique in a given segment of a dataset, over a larger timescale or greater geographic area there is ambiguity as to the identity of the user as the probability of collision increases.

\cite{demir2017pitfalls} provides a detailed explanation of the Birthday Paradox and notes an approximation to identify how many $m$ messages are approximately required for there to be at least 1 collision with $p$ probability given $n$ possible digests, this is shown in Eq. \eqref{eq:birthdayapproxoriginal}. 
This indicates that with a digest of 30 bits (one less than the 31 bits needed to represent the 87 OUIs that provided 50\% coverage in \cite{demir2014analysing}), an approximate 5\% probability of at least one collision is obtained after just $10,495$ unique MAC addresses and a 50\% probability after $38,581$ unique MAC addresses.

\begin{equation}
m \approx \sqrt{2n \times \ln{(\frac{1}{1-p})}}
\label{eq:birthdayapproxoriginal}
\end{equation}

We reformulate this approximation to provide the number of $n$ possible digests for a given number of messages $m$ and the probability $p$ that one digest corresponds to 2 or more input values as shown in Eq \eqref{eq:birthdayapprox}.  The number of bits required to store $n$ digest combinations can be found as $bits \geq \left\lceil log_2 n \right\rceil$.

\begin{equation}
n \approx \frac{m^2}{2}\times\frac{1}{\ln({\frac{1}{1-p}})}
\label{eq:birthdayapprox}
\end{equation}

For convenience, a set of pre-computed approximations can be found in Table \ref{table:birthdayprobabilitytable}. The approximation demonstrates that $1000$ inputs with an acceptable probability of at least one collision being $0.05$ can be represented in 24 bits and 21 bits where the maximum acceptable probability of at least one collision is $0.25$.

It is worth noting that in instances where $m > 1,312$ and the desired probability remains 5\%, the number of bits required can exceed the 24 bit size of the NIC suffix of the MAC address, which can be more problematic in situations where a small group of manufacturers produce devices in the sample set. 
It is also worth noting that digest sizes of 24 bits or less still appear to be suitable when the maximum desired probability is 25\% where $m \leq 3,106$ and $m \leq 4,822$ where the probability is raised to 50\%.

\begin{table}[htbp]
\caption{Approximate number of bits required in the output of a hash function for $\leq p$ probability of there being at least 1 collision with $m$ inputs.}
\begin{center}
\scriptsize
\begin{tabular}{l | llllll}
          		& $p \lessapprox 0.05$		& $p \lessapprox 0.25$		& $p \lessapprox 0.5$     & $p \lessapprox 0.75$    \\ \hline
$m \approx 100$       & 17 bits 		& 15 bits 		& 13 bits 		& 12 bits \\
$m \approx 1,000$     & 24 bits 		& 21 bits 		& 20 bits 		& 19 bits \\
$m \approx 10,000$    & 30 bits 		& 28 bits 		& 27 bits 		& 26 bits \\
$m \approx 100,000$   & 37 bits 		& 35 bits 		& 33 bits 		& 32 bits \\
$m \approx 1,000,000$ & 44 bits 		& 41 bits 		& 40 bits 		& 39 bits
\end{tabular}
\end{center}
\label{table:birthdayprobabilitytable}
\end{table}

It is critical to note the approximation approach described in Eq. \eqref{eq:birthdayapproxoriginal} and Eq. \eqref{eq:birthdayapprox} simply calculate the probability of there being at least 1 collision but do not estimate  the rate of collisions.

A more crudely modified Birthday Paradox approximation was applied for calculating a digest size for one-to-many hash based anonymization in MAC addresses in \cite{demir2017pitfalls} where it is also noted that this can quickly lead to anonymity as the number of messages increases.  
\cite{demir2017pitfalls} contend that it is difficult to control the anonymity set size when using such approximations. As can be seen in Table \ref{table:birthdayprobabilitytable}, this approach rapidly leads to digest sizes that offer no anonymity. Recall \cite{minch2015location} mentions a deployment of 12 sensors which, on a single day on the 6th July, detected $106,629$ unique MAC addresses, which represents approximately 11.3\% of all (non-unique) detections.

We take a different approach and allow for an acceptable proportion of MAC addresses to experience  collision, instead of trying to minimise the probability of at least one collision in the anonymized dataset. 
Below we derive an expression for the rate of collisions $p$ for  $m$  messages hashed across  $n$ digests (or buckets). 
The probability that a message  collides with another, i.e. when there is only one other digest stored, can be computed as $1/n$. 
The probability that no such collision exists is therefore $1 - 1/n$. 
To find the probability that $m-1$ addresses do not collide with a given digest, we can compute $(1 - 1/n)^{m - 1}$. 
Simply subtracting such a result from $1$ will therefore provide the probability that $a\,\, message$ will have a collision, given a total of $m$ messages and a digest size of $n$:

\begin{equation}
p = 1-(1-1/n)^{m-1}
\label{eq:collisionprobability}
\end{equation}

The expected number of collisions within $m$ messages is thus $C = p\times m$. 
Eq. \eqref{eq:collisionprobability} provides that with a 24 bit digest ($m = 2^{24}$), the rate $p$ exceeds 0.01 when $m > 168617$. 
Where $m \leq 2^{24}$, we are able to provide at least, or better, anonymization as the approach detailed in \cite{fuxjaeger2016towards} of removing the the OUI manufacturer prefix from MAC address. However; due to the hashing approach, we similarly are able to mitigate risks of attacks described in \cite{martin2016decomposition} due to the NIC suffix of the MAC address being allocated on a contiguous basis. 
A computed table of minimum digest sizes required to provide for a maximum $p$ rate of collision given $m$ MAC addresses is provided in Table \ref{table:probabilitytable}.

In the context of MAC addresses, the value of $m$ should be set appropriate for the detection volume over a given time period and the value $p$ should be set to the desired rate of collision. 
For example, with $10,000$ unique detections between two points in a 40 minute interval (250 vehicles per minute) and acceptable collision rate of 1\%, a hash digest of 20 bits may be suitable. 
However, with $360,000$ unique detections over a 24 hour period, assuming a consistent detection rate of unique MAC addresses, the  collision rate rises to 29.1\%.

Note that the volume of MAC addresses collected may also vary by sampling rate. 
A study in Delaware \cite{sharifi2011analysis} found the detection rate for Bluetooth traffic sensors to vary by hour from a minimum of $4.3\%$ to a maximum of $6.9\%$. This corresponded to a sampling rate (for a journey time calculation) between $2.02\%$ and $5.39\%$.

\begin{table}[htbp]
\caption{Minimum number of bits required in the output of a hash function for $\leq p$ rate of collisions in the dataset with $m$ inputs.}
\begin{center}
\scriptsize
\begin{tabular}{l | llllll}
          		& $p \leq 0.01$		& $p \leq 0.05$		& $p \leq 0.5$     & $p \leq 0.75$    \\ \hline
$m = 100$       & 14 bits 			& 9 bits 			& 8 bits 			& 7 bits \\
$m = 1,000$     & 17 bits 			& 15 bits 			& 11 bits 			& 10 bits \\
$m = 10,000$    & 20 bits 			& 18 bits 			& 14 bits 			& 13 bits \\
$m = 100,000$   & 24 bits 			& 21 bits 			& 33 bits 			& 18 bits \\
$m = 1,000,000$ & 27 bits 			& 25 bits 			& 21 bits 			& 20 bits
\end{tabular}
\end{center}
\label{table:probabilitytable}
\end{table}

\section{Evaluation}

\subsection{Experiment setup}
The goal of evaluation is to validate the expression for collision rate using realistic MAC addresses as well as obtain hash configurations where the rate of collision remains less than or equal to 1\% at intervals of $100$, $1,000$ and $10,000$ unique MAC addresses. 
We formulate an experiment to gain empirical metrics on the rate of collision for given number of MAC addresses and a given digest size. 
Such hashes are anonymized in the manner aforesaid in the previous section; the MAC address is concatenated with a randomly generated salt, a hash is computed and truncated to obtain anonymity.

We randomly generate $m$ unique MAC addresses in 00:16:3e:00:00:00 and 00:16:3e:7f:ff:ff range ($\approx 8.4 \times 10^6$ possible values), using a script recommended by RedHat for generating randomised MAC addresses for virtual machines \cite{redhat}. 
Such MAC addresses are then hashed together with a randomly-generated static 68-bit salt. 
The hash is then truncated to $log_2 n$ bits, where $n$ is the maximum number of digests that can be stored in a given hash. 
As a MAC address is hashed, in the event the truncated digest collides with a previous value, the number of collisions is incremented. 
A rate of collisions is calculated as the number of collisions divided by $m$.

This experiment is repeated $100$ times on each quantity of MAC addresses, for each digest size. On each round a new set of random MAC addresses is devised and a new random salt generated. 
The median value is recorded once all the rounds are completed.
The experiment was run on digest sizes from $2^{13}$ to $2^{21}$ and quantities of MAC addresses from $m = 100$ to $m = 100,000$. 
Due to the high amount of hashes that needed to be computed for this experiment, it was run overnight on a cloud computing environment.

\subsection{Results and Discussion}

Table \ref{table:probabilitytablecollissions} shows the results of these experiments. 
The experiment demonstrates that for $100$ MAC addresses, a digest size of 13 bits yields $1\%$ collisions but rises to $11.1\%$ in a dataset of $1,000$ hashes, $62.6\%$ in a dataset of $10,000$ hashes and $95.9\%$ in a dataset of $100,000$ hashes. 
Similarly, a digest size of $2^{17}$ is suitable for acceptable $0.7\%$ collisions in a dataset of $10,000$ hashes but increases to $7.1\%$ in $10,000$ addresses and $48.7\%$ in $100,000$ addresses. 

The experimental results closely match the approximation given by an analytical approximation. 
For datasets of $1,000$ and $10,000$ MAC addresses both the probabilistic calculation and the experimental results show digest sizes of 17 bits and 20 yield less than 1\% collisions. 
For datasets of $100$ MAC addresses, whilst the probabilistic calculation showed 13 bits would yield a 1.2\% rate of collisions, we experimentally found that 13 bits delivered the acceptable 1\% rate of collision. As the dataset was only formed of $100$ MAC Addresses, the collision rate increments by 1\% for each collision.
As a comparison, Eq \eqref{eq:collisionprobability}  indicates that for $100$ MAC addresses and a digest size of $2^{14}$, the rate of collision should be 0.6\%.
For $1,000$ MAC addresses, a digest size of $2^{17}$ should yield a rate of collisions of 0.8\%. Finally, for $10,000$ MAC addresses, a digest size of $2^{20}$ results in a rate of 1.0\%.

We accordingly demonstrate that the obtained expression is practically suitable for determining the probability of collision in a dataset of hashes, given the size of a hash digest and the number of records stored. The increase in the number of records yields a higher rate of collisions, demonstrating the anonymization effect.

\begin{table}[htbp]
\caption{Median \% of collisions (rounded to 1 decimal place) for $m$ MAC addresses hashed into $n$ buckets.}
\begin{center}
\scriptsize
\begin{tabular}{l | llllll}
				& $m = 100$	& $m = 1,000$	& $m = 10,000$	& $m = 100,000$\\ \hline
$n = 2^{13}$	& $1.0\%$	& $11.1\%$		& $62.6\%$		& $95.9\%$\\
$n = 2^{14}$	& $0.0\%$	& $5.7\%$		& $42.2\%$		& $91.8\%$\\
$n = 2^{15}$	& $0.0\%$	& $3.0\%$		& $25.2\%$		& $83.7\%$\\
$n = 2^{16}$	& $0.0\%$	& $1.4\%$		& $13.8\%$		& $68.7\%$\\
$n = 2^{17}$	& $0.0\%$	& $0.7\%$		& $7.1\%$		& $48.7\%$\\
$n = 2^{18}$	& $0.0\%$	& $0.4\%$		& $3.7\%$		& $30.0\%$\\
$n = 2^{19}$	& $0.0\%$	& $0.2\%$		& $1.9\%$		& $16.9\%$\\
$n = 2^{20}$	& $0.0\%$	& $0.1\%$		& $1.0\%$		& $9.0\%$\\
$n = 2^{21}$	& $0.0\%$	& $0.0\%$		& $0.5\%$		& $4.6\%$\\
\end{tabular}
\end{center}
\label{table:probabilitytablecollissions}
\end{table}

\section{Conclusion and Future work}

In this paper we study MAC address anonymization strategies  for Bluetooth and Wi-Fi based tracking systems. Existing approaches to anonymization using hash-based approaches have been found to be easily invertible due to a small search space and  limited number of well-known  OUI manufacturer prefixes which makes   frequency-based attacks possible. 
Whilst MAC addresses are 48 bits in length, we find that the entire allocated space of MAC addresses can be represented in no more than 39 bits and 50\% of the most common manufacturers can be represented in just 31 bits.

Contrary to prior work on anonymization of hash-based datasets that  sought to minimise the probability of "at least one" collision, we have instead formulated the problem in terms of defining a tolerable rate of collisions in the overall dataset. 
We present a practical approach to anonymizing MAC addresses by hashing MAC addresses using computationally expensive algorithms and truncating the digests to achieve $k$-anonymity. We provide an approach for calculating the rate of collisions in a dataset and experimentally demonstrate this approach in practice, showing it to be suitable to obtain hash configurations with an acceptable rate of collisions below 1\% (for example, using a digest size of 20 bits for $10,000$ MAC addresses).
 
The proposed approach opens the door for hash-based anonymization to be applied to low-entropy datasets, which is a potential further work. 
Other areas for potential investigation include an investigation of privacy-preserving opt-out mechanisms. Our work also opens the door for novel approaches in anonymous {\it contact tracing} during disease pandemics without the need to deploy new software to user devices.
Further research on de-anonymization attacks of the approach we have outlined would be of interest, particularly amongst practical datasets. 
Finally, multi-dimensional de-anonymization attacks would prove to be interesting, where such trackers capture additional information such as  walking or cycling speed, or device characteristics. 

\bibliographystyle{apalike}
{\small
\bibliography{references.bib}}

\begin{thebibliography}{}

\bibitem[Abberley et~al., 2017]{abberley2017modelling}
Abberley, L., Gould, N., Crockett, K., and Cheng, J. (2017).
\newblock Modelling road congestion using ontologies for big data analytics in
  smart cities.
\newblock In {\em 2017 International Smart Cities Conference (ISC2)}, pages
  1--6. IEEE.

\bibitem[Abedi et~al., 2015]{abedi2015assessment}
Abedi, N., Bhaskar, A., Chung, E., and Miska, M. (2015).
\newblock Assessment of antenna characteristic effects on pedestrian and
  cyclists travel-time estimation based on bluetooth and wifi mac addresses.
\newblock {\em Transportation Research Part C: Emerging Technologies},
  60:124--141.

\bibitem[Ali, 2017]{ali2017mechanism}
Ali, J. (2017).
\newblock Mechanism for the prevention of password reuse through anonymized
  hashes.
\newblock {\em PeerJ PrePrints}, 5:e3322v1.

\bibitem[Alimpia et~al., 2018]{alimpia2018enhanced}
Alimpia, J.~B., Sison, A.~M., and Medina, R.~P. (2018).
\newblock An enhanced hash-based message authentication code using bcrypt.
\newblock {\em Proceedings of the International Journal for Research in Applied
  Science and Engineering Technology}, 6(4).

\bibitem[Biryukov et~al., 2016]{biryukov2016argon2}
Biryukov, A., Dinu, D., and Khovratovich, D. (2016).
\newblock Argon2: new generation of memory-hard functions for password hashing
  and other applications.
\newblock In {\em 2016 IEEE European Symposium on Security and Privacy
  (EuroS\&P)}, pages 292--302. IEEE.

\bibitem[Blocki et~al., 2018]{blocki2018economics}
Blocki, J., Harsha, B., and Zhou, S. (2018).
\newblock On the economics of offline password cracking.
\newblock In {\em 2018 IEEE Symposium on Security and Privacy (SP)}, pages
  853--871. IEEE.

\bibitem[Blogg et~al., 2010]{blogg2010travel}
Blogg, M., Semler, C., Hingorani, M., and Troutbeck, R. (2010).
\newblock Travel time and origin-destination data collection using bluetooth
  mac address readers.
\newblock In {\em Australasian transport research forum}, volume~36.

\bibitem[{Clearview Intelligence Ltd}, 2020]{clearview}
{Clearview Intelligence Ltd} (2020).
\newblock Product specification m830 traffic monitoring.

\bibitem[Cragg, 2013]{cragg2013bluetooth}
Cragg, S. (2013).
\newblock Bluetooth detection--cheap but challenging.
\newblock In {\em Scottish Transport Applications and Research Conference
  (STAR)}.

\bibitem[Demir et~al., 2014]{demir2014analysing}
Demir, L., Cunche, M., and Lauradoux, C. (2014).
\newblock Analysing the privacy policies of wi-fi trackers.
\newblock In {\em Proceedings of the 2014 workshop on physical analytics},
  pages 39--44.

\bibitem[Demir et~al., 2017]{demir2017pitfalls}
Demir, L., Kumar, A., Cunche, M., and Lauradoux, C. (2017).
\newblock The pitfalls of hashing for privacy.
\newblock {\em IEEE Communications Surveys \& Tutorials}, 20(1):551--565.

\bibitem[Ertaul et~al., 2016]{ertaul2016implementation}
Ertaul, L., Kaur, M., and Gudise, V. A. K.~R. (2016).
\newblock Implementation and performance analysis of pbkdf2, bcrypt, scrypt
  algorithms.
\newblock In {\em Proceedings of the International Conference on Wireless
  Networks (ICWN)}, page~66. The Steering Committee of The World Congress in
  Computer Science, Computer~….

\bibitem[Fearn, 2019]{fearn_2019}
Fearn, N. (2019).
\newblock Privacy warning over tfl plans to track tube passengers via wifi |
  computing.

\bibitem[Fuxjaeger et~al., 2016]{fuxjaeger2016towards}
Fuxjaeger, P., Ruehrup, S., Paulin, T., and Rainer, B. (2016).
\newblock Towards privacy-preserving wi-fi monitoring for road traffic
  analysis.
\newblock {\em IEEE Intelligent Transportation Systems Magazine}, 8(3):63--74.

\bibitem[Hidayata et~al., 2018]{hidayata2018time}
Hidayata, A., Terabea, S., and Yaginumaa, H. (2018).
\newblock Time travel estimations using mac addresses of bus, passengers: A
  point to path-qgis analysis.
\newblock {\em Geoplanning: Journal of Geomatics and Planning}, 5(2):259--268.

\bibitem[Li et~al., 2019]{li2019protocols}
Li, L., Pal, B., Ali, J., Sullivan, N., Chatterjee, R., and Ristenpart, T.
  (2019).
\newblock Protocols for checking compromised credentials.
\newblock In {\em Proceedings of the 2019 ACM SIGSAC Conference on Computer and
  Communications Security}, pages 1387--1403.

\bibitem[Martin et~al., 2017]{martin2017study}
Martin, J., Mayberry, T., Donahue, C., Foppe, L., Brown, L., Riggins, C., Rye,
  E.~C., and Brown, D. (2017).
\newblock A study of mac address randomization in mobile devices and when it
  fails.
\newblock {\em Proceedings on Privacy Enhancing Technologies},
  2017(4):365--383.

\bibitem[Martin et~al., 2016]{martin2016decomposition}
Martin, J., Rye, E., and Beverly, R. (2016).
\newblock Decomposition of mac address structure for granular device inference.
\newblock In {\em Proceedings of the 32nd Annual Conference on Computer
  Security Applications}, pages 78--88.

\bibitem[Marx et~al., 2018]{marx2018hashing}
Marx, M., Zimmer, E., Mueller, T., Blochberger, M., and Federrath, H. (2018).
\newblock Hashing of personally identifiable information is not sufficient.
\newblock {\em SICHERHEIT 2018}.

\bibitem[Minch, 2015]{minch2015location}
Minch, R.~P. (2015).
\newblock Location privacy in the era of the internet of things and big data
  analytics.
\newblock In {\em 2015 48th Hawaii International Conference on System
  Sciences}, pages 1521--1530. IEEE.

\bibitem[{Red Hat, Inc.}, 2020]{redhat}
{Red Hat, Inc.} (2020).
\newblock Generating a new unique mac address.

\bibitem[Sharifi et~al., 2011]{sharifi2011analysis}
Sharifi, E., Hamedi, M., Haghani, A., and Sadrsadat, H. (2011).
\newblock Analysis of vehicle detection rate for bluetooth traffic sensors: A
  case study in maryland and delaware.
\newblock In {\em 18th World Congress on on Intelligent Transport Systems}.

\bibitem[Thomas et~al., 2019]{thomas2019protecting}
Thomas, K., Pullman, J., Yeo, K., Raghunathan, A., Kelley, P.~G., Invernizzi,
  L., Benko, B., Pietraszek, T., Patel, S., Boneh, D., et~al. (2019).
\newblock Protecting accounts from credential stuffing with password breach
  alerting.
\newblock In {\em 28th USENIX Security Symposium USENIX Security 19)}, pages
  1556--1571.

\bibitem[Velisavljevic et~al., 2016]{vladan2016magnetic}
Velisavljevic, V., Cano, E., Dyo, V., and Allen, B. (2016).
\newblock Wireless magnetic sensor network for road traffic monitoring and
  vehicle classification.
\newblock {\em Transport and Telecommunication Journal}, 17(4).

\bibitem[Wetzels, 2016]{wetzels2016open}
Wetzels, J. (2016).
\newblock Open sesame: The password hashing competition and argon2.
\newblock {\em arXiv preprint arXiv:1602.03097}.

\end{thebibliography}

\end{document}